\documentclass[prl,amsmath,twocolumn,showpacs,superscriptaddress]{revtex4}

\usepackage{graphicx}
\usepackage{color}
\begin{document}

\title{Superdiffusive dispersals impart the geometry of underlying random walks  
}
\author{V. Zaburdaev}
\affiliation{Max Planck Institute for the Physics of Complex Systems,~N\"{o}thnitzer Strasse 38, D-01187 Dresden, Germany}
\affiliation{Institute of Supercomputing Technologies, Lobachevsky State University of Nizhny Novgorod, 603140 N. Novgorod, Russia}
\author{I. Fouxon}
\affiliation{Department of Physics, Institute of Nanotechnology and Advanced Materials, Bar-Ilan University, Ramat-Gan,
52900, Israel}
\author{S. Denisov}
\affiliation{Department of Applied Mathematics, Lobachevsky State University of Nizhny Novgorod, 603140 N. Novgorod, Russia}
\affiliation{Sumy State University, Rimsky-Korsakov Street 2, 40007 Sumy, Ukraine}
\affiliation{Institute  of Physics, University of Augsburg,
Universit\"atsstrasse 1, D-86135,  Augsburg 
Germany}
\author{E. Barkai}
\affiliation{Department of Physics, Institute of Nanotechnology and Advanced Materials, Bar-Ilan University, Ramat-Gan, 
52900, Israel}

\begin{abstract}

{It is recognised now that a variety of real-life phenomena ranging from diffuson of 
cold atoms to motion of humans exhibit dispersal faster than 
normal diffusion. L\'{e}vy walks is a  model that excelled 
in describing such superdiffusive behaviors albeit in one dimension. 
Here we show that, in contrast to standard random walks, the microscopic 
geometry of planar superdiffusive L\'{e}vy walks is imprinted 
in the asymptotic distribution of the walkers. 
The geometry of the underlying walk can be inferred from  trajectories of the walkers 
by calculating the analogue of the Pearson coefficient.
}
\end{abstract}

\maketitle

\textit{Introduction.} The {L}\'{e}vy walk (LW) model \cite{yosi1,yosi2,rmp} was developed to describe
spreading  phenomena that were not fitting 
the  paradigm of Brownian diffusion \cite{beyond}. Still looking as a random walk, see Fig.~1,
but with a very broad distribution of excursions' lengths, the corresponding processes 
exhibit dispersal faster than in the case of normal diffusion. 
Conventionally, this difference is quantified with the mean squared displacement (MSD), 
$\langle r^2(t) \rangle \propto t^{\alpha}$, and the regime with $\alpha > 1$ 
is called super-diffusion. Examples of such systems  range from cold atoms 
moving in dissipative optical lattices \cite{sagi2012} to T cells migrating in the brain tissue \cite{Harris2012}. 
Most of the existing theoretical results, however,
were derived for one dimensional LW processes \cite{rmp}. 
In contrast, real life phenomena -- biological motility (from bacteria \cite{taktikos}  to humans \cite{gadza}
and autonomous robots \cite{robot1,swarms}), 
animal foraging \cite{mendes,foraging} and search \cite{search} 
--  happen in two dimensions. Somewhat surprisingly, generalizations of the L\'{e}vy walks
to two dimensions are still virtually unexplored.

In this work we extend the concept of LWs to two dimensions. 
Our main finding is that the microscopic geometry of planar L\'{e}vy walks
reveals itself in the shape of  the asymptotic probability density 
functions (PDF) $P(\textbf{r},t)$ of finding a 
particle at position $\textbf{r}$ at time $t$ after it was launched from the origin.
This is in a sharp contrast to the standard 2D random walks, where, by virtue of 
the central limit theorem (CLT) \cite{gnedenko}, the asymptotic PDFs do not depend on geometry of the walks and
have a universal form of the two-dimensional Gaussian distribution \cite{spitzer,sokolov}.

\begin{figure}[t]
\includegraphics[width=0.4\textwidth]{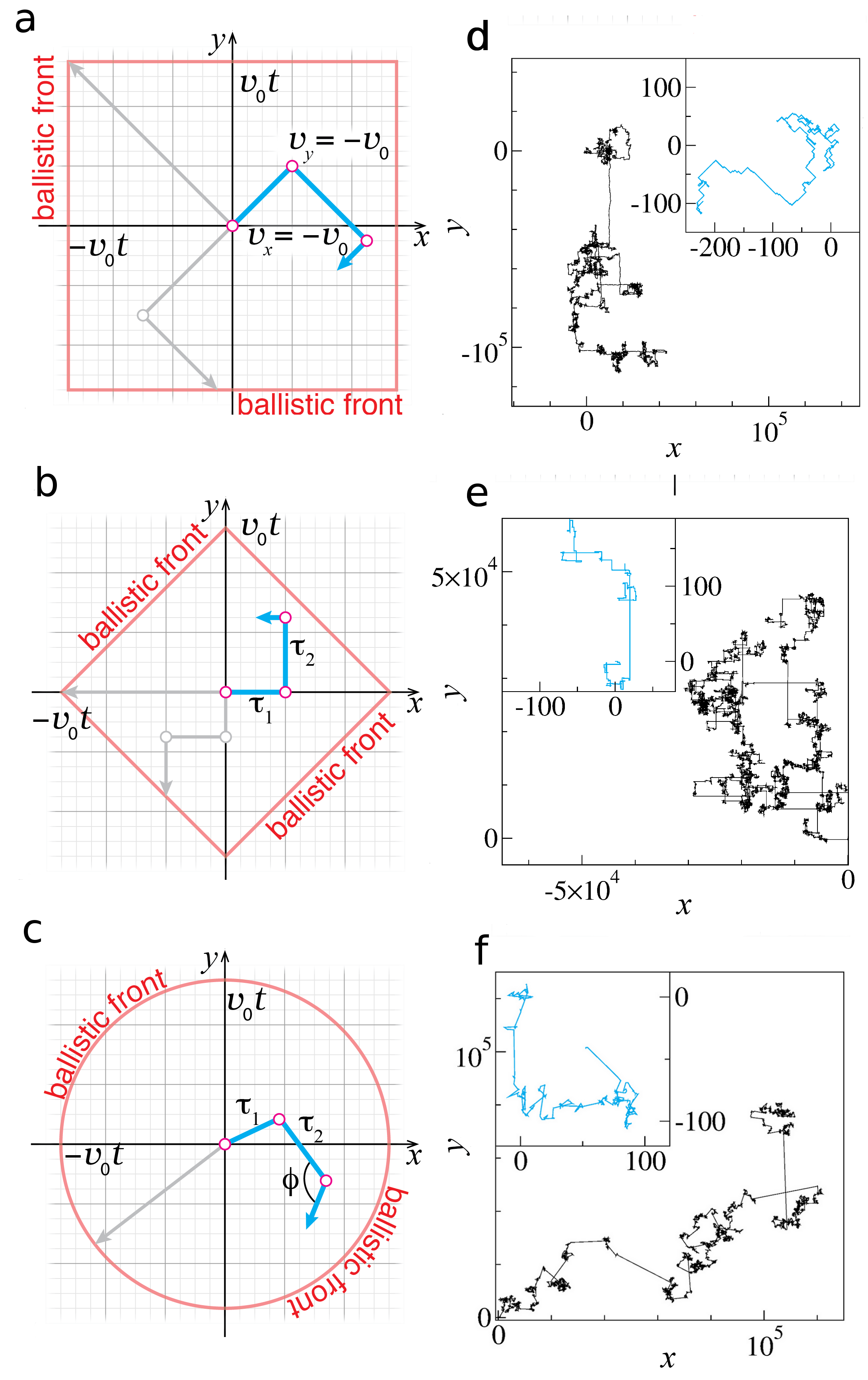}
\caption{\textbf{Three models of L\'{e}vy walks on a plane} a) In the \textit{product model} $x$ and $y$ coordinates
of a particle change according to two independent 1D L\'{e}vy walks along the coordinate axes. Whenever a direction of motion of one of the two LWs changes,
there is a kink in the trajectory ($\circ$). The ballistic front is given by
$|x|=|y|=v_0t$ (red line). b) In  \textit{$XY$-model} a particle is allowed 
to move with a speed $v_0$ only along one axis at a time which is chosen randomly at the re-orientation points $\circ$. The ballistic front is specified by the condition  $|x|+|y|=v_0t$. 
c) In the \textit{uniform model}, 
at each re-orientation point $\circ$, a particle chooses 
a random direction of motion, specified by an angle $\phi$ uniformly distributed in the interval $[0,2\pi]$,
and then moves with a constant speed $v_0$. 
The ballistic front is a circle of the radius $v_0t$.
e-f) Trajectories produced by the models (a-b) after time $t=10^6$. 
Note that on the large time scale the trajectories of the product and $XY$-models appear to be similar.
The parameters are $\gamma = 3/2$, $\upsilon_0 = 1$ and $\tau_0=1$.}
\label{Fig1}
\end{figure}

\textit{Models.} We begin with a core of the L\'{e}vy walk concept \cite{yosi1,yosi2}: A particle 
performs  ballistic moves with constant speed, alternated by 
instantaneous re-orientation events,
and the length of the moves is a random variable with a power-law distribution. 
Because of the constant speed $v_0$, the length $l_i$ and duration $\tau_i$
of the $i$-th move are linearly coupled, $l_i=v_0\tau_i$. As a result, the model can be equally 
well defined by the distribution of ballistic move (flight) times
\begin{equation}
\psi(\tau)=\frac{1}{\tau_0}\frac{\gamma}{(1+\tau/\tau_0)^{1+\gamma}},\quad \tau_0,\gamma>0. \label{fPDF}
\end{equation} 
Depending on the value of $\gamma$, it can 
lead to a dispersal $\alpha=1$, 
typical for normal diffusion ($\gamma > 2$), and 
very long excursions leading to the fast dispersal with $1 < \alpha \leq 2$ in the case of super-diffusion ($0 < \gamma < 2$). At each moment of time $t$ 
the finite speed sets the  ballistic front beyond which there 
are no particles. 
Below we consider three intuitive models of two-dimensional superdiffusive dispersals.



a) The simplest way to obtain  two-dimensional 
L\'{e}vy walk out of the one-dimensional one is to assume that
the motions along each axis, $x$ and $y$, are 
identical and independent one-dimensional LW processes, as shown in Fig~1a. 
The two-dimensional PDF  $P(\mathbf{r},t)$, $\mathbf{r}(t)=\{x(t),y(t)\}$,  of this \textit{product model}
is given by the product of two one-dimensional LW PDFs,
$P_{\text{prod}}(\mathbf{r},t)=P_{\text{LW}}(x,t)\cdot P_{\text{LW}}(y,t)$.
On the microscopic scale, each ballistic event corresponds to the motion along either the 
diagonal or anti-diagonal. 
Every re-orientation only partially erases the memory about the last ballistic flight: 
while the direction of the motion 
along one axis could be changed, the direction along the other axis almost surely remains the same. 
The ballistic front  has the shape of a square with borders given by $|x|=|y|=v_0t$.

b) In the \textit{XY-model}, a particle is allowed to 
move only along one of the axes at a time. 
A particle chooses a random flight time $\tau$ from Eq. (\ref{fPDF})
and one out of four directions. 
Then it moves with a constant speed $\upsilon_0$ along the chosen direction. 
After the flight time is elapsed, a new random direction and a new flight time are chosen. This process is sketched in Fig. 1b. 
The ballistic front is a square defined by the equation $|x|+|y|=v_0t$.


c) The \textit{uniform  model} follows the  original definition by Pearson \cite{Pearson1905}. 
A particle chooses a random direction, 
parametrized by the angle $\phi$, uniformly distributed in the interval $[0,2\pi]$, 
and then moves ballistically  for a chosen flight time. The direction of the next flight is chosen  randomly and independently of the direction of the elapsed flight.
The corresponding trajectory is sketched in Fig. 1c. The ballistic front of the model 
is a circle $|\mathbf{r}|=v_0t$.

\begin{figure*}[t]
\includegraphics[width=1.\textwidth]{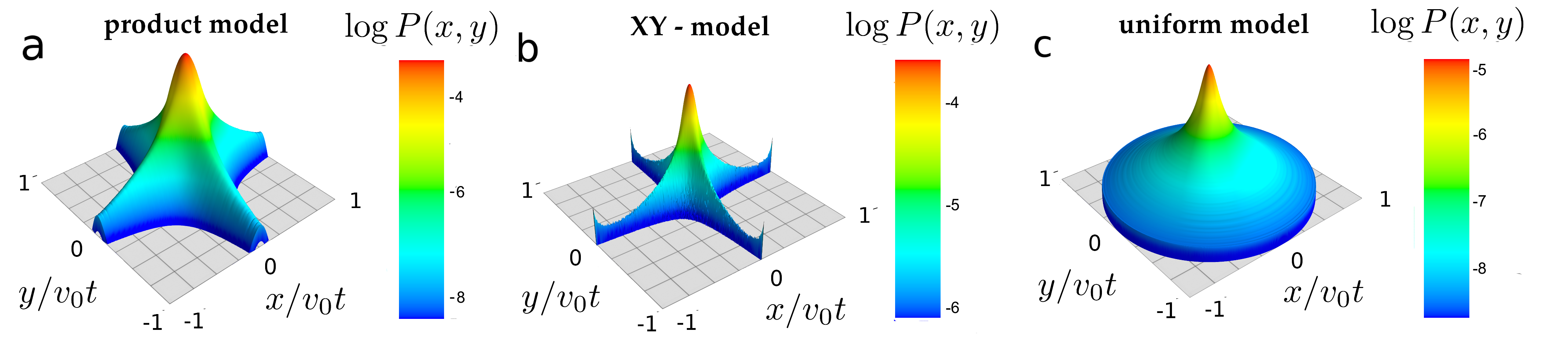}
\caption{\textbf{Probability density functions of the three models in the super-diffusive regime.} 
The distributions are plotted on a log scale for the time $t/\tau_0 = 10^4$. The PDF 
for the product model (a) was obtained by multiplying PDFs
of two identical one-dimensional LW processes.  The PDFs for the $XY$ and uniform
models were obtained by sampling over $10^{14}$  realizations.
The parameters are $\gamma = 3/2$, $v_0 = 1$ and $\tau_0=1$.}
\label{Fig2}
\end{figure*}

\textit{Governing equations.}
We now derive equations describing the density of particles for the $XY$ and uniform models.
The following two coupled equations govern the space-time evolution of the PDF \cite{rmp}:
\begin{equation}
\nu(\mathbf{r},t)=\int_{0}^{t}{\rm d}\tau\int{\rm d}\mathbf{v}\,\nu(\mathbf{r}-\mathbf{v}\tau,t-\tau)\psi(\tau)h(\mathbf{v})
+\delta(\mathbf{r})\delta(t),\nonumber
\end{equation}
\begin{equation}
P(\mathbf{r},t)=\int_{0}^{t}{\rm d}\tau\int{\rm d}\mathbf{v}\,\nu(\mathbf{r}-\mathbf{v}\tau,t-\tau)\Psi(\tau)h(\mathbf{v}).
\label{P}
\end{equation}
The first equation describes the events of velocity reorientation [marked by $\circ$ in Fig. 1(b-c)], 
where the density $\nu(\mathbf{r},t)$ allows us to count  the number of particles, $\nu(\mathbf{r},t)d\mathbf{r}dt$,
whose flights ended in the interval $[\mathbf{r}, \mathbf{r}+d\mathbf{r}]$ during the time interval
$[t, t+dt]$. Velocity at each reorientation event is chosen from the model specific velocity distribution $h(\mathbf{v})$ and 
is statistically independent of the flight time. Second equation relates the 
events of velocity reorientations to the density of the particles. Here
$\Psi(\tau)$ is the probability to remain in the flight for time $\tau$, 
$\Psi(\tau)=1-\int_0^\tau\psi(t')\rm{d}t'$.  The formal solution of the transport equations can 
be found via combined Fourier-Laplace transform \cite{si}:
\begin{equation}
P(\mathbf{k},s)=\frac{\int {\rm d}\mathbf{v}\,\Psi(s+i\mathbf{k}\cdot\mathbf{v})h(\mathbf{v})}{1-\int {\rm d}\mathbf{v}\,\psi(s+i\mathbf{k}\cdot\mathbf{v})h(\mathbf{v})}
\label{Pks}
\end{equation}
This is a general answer for a random walk process in arbitrary dimensions 
with an arbitrary velocity distribution, where $\mathbf{k}$ and $s$ are coordinates in Fourier and Laplace space corresponding to $\mathbf{r}$ and $t$ respectively (but not for the product model, which is described by {\em two} independent random walk processes). The microscopic geometry of the process can 
be captured with $h(\mathbf{v})$. For the $XY$-model we have $h_{XY}(\mathbf{v}) = [\delta(v_y)\delta(|v_x|-v_0) + \delta(v_x)\delta(|v_y|-v_0)]/4$, 
while for the uniform model it is $h_{\text{uniform}}(\mathbf{v})=\delta(|\mathbf{v}|-v_0)/2\pi v_0$. 
The technical difficulty is to find the inverse transform of Eq.~(\ref{Pks}). We therefore employ 
the asymptotic analysis \cite{yosi1,yosi2,rmp}
to switch from Fourier-Laplace representation to the space-time coordinates
and 
analyze model PDFs $P(\mathbf{r},t)$ in the limit of large $\bf{r}$ and $t$ \cite{si}.

In the diffusion limit, $\gamma > 2$, the mean squared flight length  is
finite.  In the large time limit, the normalized covariance of the flight components in all three models
is the identity matrix and so the cores of their PDFs are governed by
the vector-valued CLT \cite{van} and have the universal
Gaussian shape $P(\mathbf{r},t)\simeq \frac{1}{4\pi D t}{\rm e}^{-\frac{\mathbf{r}^2}{4Dt}}$, 
where $\quad D=v_0^2\tau_0/[2(\gamma-2)]$ (for the product model the velocity has to be rescaled to $v_0/\sqrt{2}$). 
For the outer parts of the PDFs some bounds can be obtained based on a theory developed   
for sums of random variables with slowly decaying regular distributions \cite{borovkov}.

The difference between the three walks becomes sharp
in the regime of sub-ballistic super-diffusion, $1 < \gamma < 2$.
Figure 2 presents the PDFs of the three models obtained by  sampling \cite{si} over the corresponding 
stochastic processes for  $t =10^4 \gg \tau_0 = 1$. 
These results reveal a striking feature, namely, that the geometry of 
a random walk is imprinted in its PDF. This is very visual close
to the ballistic fronts, however, as we show below, the non universality is already 
present in the PDF cores.


The  PDF of the product model is the product of the PDFs for two identical
one-dimensional LWs \cite{rmp}. In the case of the $XY$-model, 
the central part of the propagator can be written in Fourier-Laplace space as
$P_{XY}(k_x,k_y,s)\simeq (s+\frac{K_{\gamma}}{2}|k_x|^{\gamma}+\frac{K_{\gamma}}{2}|k_y|^{\gamma})^{-1}$,
where $K_{\gamma}=\Gamma[2-\gamma]|\cos(\pi\gamma/2)|\tau_0^{\gamma-1}v_0^{\gamma}$ \cite{si}. 
By inverting the Laplace transform, we also arrive at the product of two characteristic 
functions of one-dimensional L\'{e}vy distributions \cite{uchaikinzolotarev,span1}:
$P_{\text{XY}}(k_x,k_y,t)\simeq\text{e}^{-tK_{\gamma}|k_x|^{\gamma}/2}\text{e}^{-tK_{\gamma}|k_y|^{\gamma}/2}$.
In this case the spreading of 
the particles along each axis happens twice slower (note a factor $1/2$ in the exponent) 
than in the one-dimensional case; each excursion along an axis acts as a trap for the motion along 
the adjacent axis thus 
reducing the characteristic time of the dispersal process  by factor $2$. 
As a result, the bulk PDF of the $XY$-model is similar to 
that of the product model after the velocity rescaling $\tilde{v}_0 = v_0/2^{1/\gamma}$. This explains why on the macroscopic scales the trajectory of the product model, see Fig.~1e, looks
similar to that of the $XY$ - model. The difference between the PDFs of these two models appears
in the outer parts of the distributions, see Figs.~2a,b; it
can not be resolved  with the asymptotic analysis which addresses only 
the central cores of the PDFs. The 
PDF of the $XY$-model has a cross-like structure with sharp peaks at the ends, see Fig.~3a. 
The appearance of these Gothic-like
`flying buttresses' \cite{stone}, capped with `pinnacles', can be understood by analyzing 
the process near the ballistic fronts \cite{si}.

For the uniform model we obtain 
$P_{\text{uniform}}(\mathbf{r},t) \simeq 
\frac{1}{2\pi}\int\limits_{0}^{\infty}J_{0}(kr)\text{e}^{-t\widetilde{K}_{\gamma}k^{\gamma}}k{\rm d}k$,
where 
$\widetilde{K}_{\gamma}=\tau_0^{\gamma-1}v_0^{\gamma}\sqrt{\pi}\Gamma[2-\gamma]/\Gamma\left[1+\gamma/2\right]\left|\Gamma\left[(1-\gamma)/2\right]\right|$,
and $J_{0}(x)$ is the Bessel function of the first kind (see \cite{si} for more details).  
Different to the product and $XY$-models, 
this  is a radially symmetric function which naturally follows from the microscopic isotropy of the  walk.  
Mathematically, the expression above is a  generalization of the L\'{e}vy distribution 
to two dimensions \cite{uchaikinzolotarev,klages}. However, from the physics point of view, it 
provides the generalization of the Einstein relation and relates the generalized 
diffusion constant $\widetilde{K}_{\gamma}$ to the physical parameters of the 2d process, 
$v_0$, $\tau_0$ and $\gamma$. In Fig. 3b  we compare the simulation 
results for the PDF of the uniform model with the analytical expression above.

The regime of ballistic diffusion occurs when the mean flight time diverges,  $0<\gamma<1$ \cite{daniela,madrigas}. 
Long flights dominate the distribution of particles and this
causes the probability concentration at the ballistic fronts.
Since the latter are model specific, see Fig.~1, the difference in the microscopic schematization reveals 
itself in the  PDFs even more clearly \cite{si}.



\textit{Pearson coefficient.}
The difference in the model PDFs can by quantified  by looking into  moments of the corresponding processes. The most common
is the MSD,  
$\langle\mathbf{r}^2(t)\rangle = \int{\rm d}\mathbf{r}\,\,\mathbf{r}^2 P(\mathbf{r},t)$.  
Remarkably,  
for the $XY$- and uniform  models the MSD 
is the same as for the 1D L\'{e}vy walk with the distribution of  flight times given by  Eq. (1) \cite{si}. 
The MSD therefore does not differentiate between the $XY-$ and uniform random walks (and, if the 
velocity $v_0$ is not known \textit{a priori}, the product random walks as well). Next are the fourth 
order moments, including the cross-moment $\langle x^2(t)y^2(t) \rangle$. They can be evaluated 
analytically for all three models \cite{si}.
The ratio between the cross-moment and the product of the second moments,
$PC(t) = \langle x^2(t)y^2(t)\rangle/\langle x^2(t)\rangle \langle y^2(t)\rangle$,
is a scalar characteristic similar to the Pearson coefficient \cite{pears2,ken}. 
In the asymptotic limit and in the most interesting regime of sub-ballistic super-diffusion, $1 < \gamma < 2$,
this generalized Pearson coefficient equals
\begin{equation}
PC(t)\! =\!\!   \left\{\!\!
                \begin{array}{cc}
                 1, & \text{product model} \\
               
                \frac{\gamma\Gamma[4-\gamma]^2}{\Gamma[7-2\gamma]}, & XY-\text{model}\\
               
                  \frac{(2-\gamma)^2(3-\gamma)^2}{2(4-\gamma)(5-\gamma)(\gamma-1)}(\frac{t}{\tau_0})^{\gamma-1} & \text{uniform model}
                \end{array}
              \right.
              \label{pc}
\end{equation}
The $PC$ parameter is distinctly different for the three processes: the product model has the $PC(t) \equiv 1$, 
for the $XY$-model it is a constant smaller than one for any $\gamma \in~ ]1,2]$ and does not depend on $\upsilon_0$ and $\tau_0$, 
while for the uniform model it diverges in the asymptotic 
limit as $t^{\gamma-1}$. Figure 3 presents the $PC(t = 5 \cdot 10^5)$ of the $XY-$ (c) and uniform models (d) 
obtained by samplings over $10^{14}$ stochastic realizations. We attribute the deviation of the numerical 
results for the $XY$-model from the analytical result Eq.~(\ref{pc}) near $\gamma \lesssim 2$  to a slow
convergence to the asymptotic limit $PC(t \rightarrow \infty)$  \cite{si}.

\begin{figure}[t]
\includegraphics[width=0.47\textwidth]{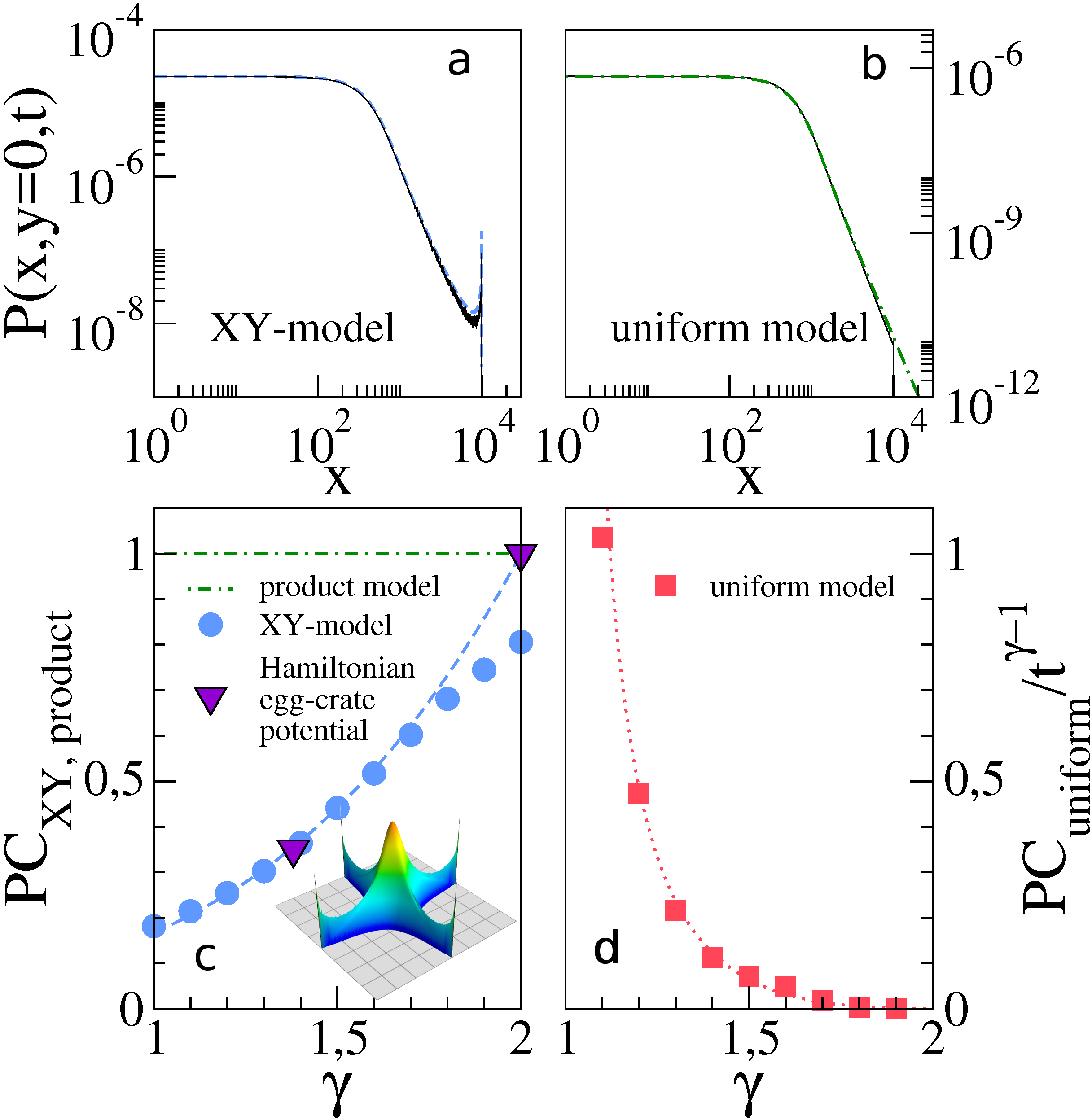}
\caption{\textbf{Statistical features of the models and their Pearson coefficients.} 
a-b) The section of the PDF of the  $XY$ model (a) and uniform model (b) 
along $x$ axis. 
The results of the statistical sampling for $t=10^4$ (solid black line)
are compared with the analytical results (dashed lines): for the 
$XY$ model it is a product of the
\textit{one}-dimensional L\'{e}vy distribution and the function $t^{1/\gamma}(t-x/v)^{-1/\gamma}$ \cite{si}, 
for the uniform model it is a \textit{two}-dimensional L\'{e}vy distribution. c-d) Pearson coefficients 
for three models. Lines correspond to the asymptotic values Eq. (4), symbols present the results of 
statistical sampling for the time $t =10^5$ (error bars are smaller then the symbol size).
The $PC$s for 
the chaotic Hamiltonian diffusion 
in an egg-crate potential \cite{egg} at time $t = 10^5$, 
for  energy values $E=4$ (left triangle) and $E=5.5$ (right triangle), was obtained by sampling
over $10^5$ independent realizations. The values of the exponent $\gamma$, $1.38$ and $2$, were extracted 
from the  MSD exponent $\alpha$, $\gamma = 3 - \alpha$.  Inset shows the PDF of the process for $t = 10^3$ 
sampled over $10^8$ realizations.
}
\label{Fig3}
\end{figure}

$PC$ can be used to find how close is a particular two-dimensional super-diffusive process
to each of the models. The value of $\gamma$ can be estimated from
the MSD exponent $\alpha$, $\gamma = 3 - \alpha$. To test this idea we investigate a classical
two-dimensional chaotic Hamiltonian system \cite{geisel,egg}
which exhibits a super-diffusive LW-like dynamics \cite{egg, beyond}. In this system, a particle moves
in a dissipationless egg-crate potential and, depending on its total energy, 
exhibits normal  or super-diffusive dispersals \cite{si}.
It is reported in Ref.~\cite{egg} that for the energy $E = 4$ the dispersal is strongly anomalous,
while in Ref.~\cite{geisel} it is stated that 
the diffusion is normal with $\alpha = 1$, within the energy range $E \in [5,6]$.
We sampled the system dynamics for two energy values, $E=4$ and $5.5$. The obtained 
MSD exponents are $1.62 \pm 0.04$ and $1 \pm 0.02$, respectively. We estimated the $PC(t)$ for the time
$t = 10^5$ and obtained values $0.35$ and $0.997$.
The analytical value of the $PC$ (4) for the $XY$-process with
$\gamma = 3 - 1.62 = 1.38$ is $0.355$. This $PC$ value thus suggests that we are 
witnessing a  super-diffusive $XY$  L\'{e}vy walk. The numerically sampled PDF of 
the process \cite{si}, see inset  in Fig.~3c, confirms this conjecture.

In contrast to the uniform model, the $PC$ parameters for the $XY$ and product models are not invariant with respect to rotations of the reference frame,
$\{x',y'\}=\{x\cos\phi-y\sin\phi, x\sin\phi+y\cos\phi\}$. 
While in theory the frame can be aligned with the directions of maximal spread exhibited by an anisotropic particle density at long times, see Fig.~2a,b,
it might be not so evident in  real-life settings. The angular dependence of the $PC$  can be 
explored by rotating the reference frame by an angle $\phi\in[0,\pi/2]$, starting from some initial 
orientation, and calculating  dependence $PC[\phi]$. The result can then be compared to analytical predictions for the asymptotic limit
where the three models show different angular dependencies \cite{si}. 
In addition, the time evolution of $PC[\phi]$ is quantitatively different for 
the product and $XY$-models and thus can be used to discriminate between the two processes. 
In the product model, the dependence 
$PC[\phi]$  changes with time qualitatively. For short times it reflects the diagonal ballistic motion of particles and for 
longer times attains the shape characteristic to the $XY$ - model \cite{si}; an effect which we could already anticipate 
from inspecting the trajectories in Fig.~1d. In the $XY$ - model the positions of minima and maxima of $PC[\phi]$ are time-independent.

\textit{Conclusion.} We have considered three intuitive models of planar L\'{e}vy
walks. Our main finding is that the geometry of a walk appears to be imprinted into
the asymptotic distributions of walking particles, both in the core of the distribution and in its tails. 
We also proposed a scalar characteristic which can be used
to differentiate between the types of walks.
Further analytical results can be obtained for arbitrary velocity distribution and dimensionality of the
problem \cite{itz}. For example, it is worthwhile to explore the connections between underling 
symmetries of 2D Hamiltonian potentials and the symmetries of the emerging LWs \cite{zas1}.

The existing body of results on two-dimensional super-diffusive  phenomena demonstrates that 
the three models we considered have potential applications. A spreading  of cold atoms in a two-dimensional
dissipative optical potential \cite{atoms} is a good candidate for a realization of the product model.
Lorentz billiards \cite{sinai,bg1990,cristadoro} reproduce  the $XY$ L\'{e}vy walk with  
exponent $\gamma=2$. The uniform model
is relevant to the problems of foraging, motility of microorganisms, 
and mobility of humans \cite{rmp, mendes, foraging, sims1,sims2}. 
On the physical side, the uniform model is relevant to a L\'{e}vy-like 
super-diffusive motion of a gold nanocluster on a  plane of graphite \cite{nanocluster}
and a graphene flake placed on a graphene sheet \cite{graphene2}. 
LWs were also shown, under certain conditions, to be the optimal strategy for searching 
random sparse targets \cite{optimalsearch,search}. The performance of searchers 
using different types of 2D LWs (for example,
under specific target arrangements) is a perspective topic \cite{benichou}. 
Finally, it would be interesting to explore a non-universal behavior of 
systems driven by different types of multi-dimensional L\'{e}vy noise \cite{span2,span3,span4}.

\begin{acknowledgments}
This work was supported by the the Russian Science Foundation grant No. 16-12-10496 (VZ and SD).
IF and EB acknowledge the support by the Israel Science Foundation.
\end{acknowledgments}


\vspace{4.ex}

\begin{widetext}

\makeatletter \renewcommand{\fnum@figure}{{\bf{\figurename~S\thefigure}}}
\setcounter{figure}{0}
\setcounter{equation}{0}
\renewcommand{\theequation}{S\arabic{equation}}

\section{Supplemental Material}

\subsection{Probability density functions  of 2D L\'{e}vy walks: general solution}
The main mathematical tool 
we use to resolve the integral transport equations is the combined 
Fourier-Laplace transform with respect to space and time, defined as:
\begin{equation}
\hat{f}(\mathbf{k},s)=\int\limits_{0}^{\infty}{\rm d} t\int{\rm d}\mathbf{r}\,\text{e}^{-i\mathbf{k}\mathbf{r}}\text{e}^{-st}f(\mathbf{r},t).
\label{FLdef}
\end{equation}
The coordinates in Fourier and Laplace spaces are $\mathbf{k}$ and $s$ 
respectively. The corresponding inverse transform is defined in the 
standard way [S1].  In the text and in the following we omit the hat sign and distinguish transformed functions by their arguments.

To find a solution to the system in Eq.~(2) in the main text, we apply Fourier-Laplace 
transform, use its shift property,
\begin{equation}
\int{\rm d}\mathbf{r} \,\text{e}^{-i\mathbf{k}\mathbf{r}} f(\mathbf{r}-\mathbf{a})=\text{e}^{-i\mathbf{k}\mathbf{a}}\hat{f}(\mathbf{k}); \quad \int\limits_{0}^{\infty}{\rm d} t\,\text{e}^{-st}\text{e}^{-at}f(t)=\hat{f}(s+a).
\label{shift}
\end{equation} and obtain:
\begin{eqnarray}
\nu(\mathbf{k},s)&=&\nu(\mathbf{k},s)\int {\rm d}\mathbf{v}\,\psi(s+i\mathbf{k}\mathbf{v})h(\mathbf{v})+1,\\
P(\mathbf{k},s)&=&\nu(\mathbf{k},s)\int {\rm d}\mathbf{v}\,\Psi(s+i\mathbf{k}\mathbf{v})h(\mathbf{v}).
\end{eqnarray}
A system of integral equations in normal coordinates and time is reduced to a system of 
two linear equations for Fourier-Laplace transformed functions; it is easily resolved to give Eq.~(3):
\begin{equation}
P(\mathbf{k},s)=\frac{\int {\rm d}\mathbf{v}\,\Psi(s+i\mathbf{k}\mathbf{v})h(\mathbf{v})}{1-\int {\rm d}\mathbf{v}\,\psi(s+i\mathbf{k}\mathbf{v})h(\mathbf{v})}
\label{PksSI}
\end{equation}
It is important to note that this answer is valid for both $XY$- and uniform models, but not for the product model. 

The technical difficulty of finding the inverse Fourier-Laplace transform is the coupled nature of the problem, where space and time enter the same argument. One way to address the problem is to do asymptotic analysis. Instead of looking at full transformed functions we may consider their expansions with respect to small $\mathbf{k},s$, corresponding to large space and time scales in the normal coordinates. 
There are two such functions in the general answer Eq.~(3) in the main text, $\psi(s+i\mathbf{k}\mathbf{v})$ and $\Psi(s+i\mathbf{k}\mathbf{v})$ (note that those are Laplace transforms only, where the Fourier coordinate $\mathbf{k}$ enters the same argument together with $s$). In fact their Laplace transforms are related $\Psi(s)=[1-\psi(s)]/s$, therefore it is sufficient to show the asymptotic expansion of $\psi(s)$ for a small argument:
\begin{eqnarray}
\psi(s+i\mathbf{k}\mathbf{v})\simeq1&-&\frac{\tau_0}{\gamma-1}(s+i\mathbf{k}\mathbf{v})-\tau_0^{\gamma}\Gamma[1-\gamma](s+i\mathbf{k}\mathbf{v})^{\gamma}\nonumber\\
&+&\frac{\tau_0^2}{(\gamma-2)(\gamma-1)}(s+i\mathbf{k}\mathbf{v})^2+... .\label{expansion}
\end{eqnarray}
Depending on the value of $\gamma$ different terms have the dominating role 
in this expansion. Such for $\gamma>2$ zeroth, first and second order terms are 
dominant, leading to the classical diffusion. In the intermediate regime $1<\gamma<2$ zeroth, 
linear and a term with a fractional power of $s+i\mathbf{k}\mathbf{v}$ are dominant and 
lead to the L\'{e}vy distribution.

Now by using this asymptotic expansions we can express 
the propagators. For example, the propagator of the uniform model in the 
sub-ballistic regime, after integration with respect to $h(\mathbf{v})$,  has the asymptotic form:
\begin{equation}
P_{\text{uniform}}(\mathbf{k},s)\simeq\frac{1}{s+|\mathbf{k}|^{\gamma}\frac{\tau_0^{\gamma-1}v_0^{\gamma}\sqrt{\pi}\Gamma[2-\gamma]}{\Gamma[1+\gamma/2]|\Gamma[(1-\gamma)/2]|}}.
\end{equation}
The inverse Laplace of this expression yields an exponential function, 
and an additional inverse Fourier transform in polar coordinates leads to the 2D  L\'{e}vy distribution 
discussed in the main text.

We can also write down a similar asymptotic result for the central part of the PDF in the sub-ballistic regime $1<\gamma<2$ for arbitrary (but symmetric) velocity distribution $h(\mathbf{v})=h(-\mathbf{v})$, for arbitrary dimension $d$:
\begin{equation}
P({\bf r}, t) \simeq \int \exp\left[ i {\bf k} \cdot {\bf r} - 
t \tilde{A} \left|
\cos \left( { \pi \gamma \over 2} \right) \right| \langle \left| 
{\bf k} \cdot {\bf v} \right|^\gamma \rangle \right]
{{\rm d} {\bf k} \over (2 \pi)^d} .
\label{eqNNUU}
\end{equation}
where $\tilde{A} = (\tau_0)^{\gamma -1} \Gamma[2-\gamma]$.
The spatial  statistics $P({\bf r} ,t)$
 is controlled by the average over the velocity PDF
via a function $\langle \left| {\bf k} \cdot {\bf v} \right|^\gamma \rangle$ 
which depends on the direction of ${\bf k}$. 
This is very different from
the Gaussian case where only the  covariance matrix  of the velocities enters
in the asymptotic limit of $P({\bf r} , t)$. In this sense the
 PDF of L\'evy
walks Eq.~(\ref{eqNNUU}) is non-universal if compared with one dimensional
L\'evy walks, or normal  $d$ dimensional diffusion. 
 
\subsection{Difference between the XY and product models in the sub-ballistic regime} The asymptotic analysis 
shows that in the bulk, the $XY$ model is identical to the product model. 
Therefore the cross-section of the $XY$ PDF along the $x$ (or $y$) axis 
is well approximated by the standard 1D L\'{e}vy distribution (see Fig. 3a,b), similarly 
to the product model. There is, however, a significant deviation at the tail close to the ballistic front. 
Let us look at the $XY$ model closer to the front. Consider the density 
on the $x$ axis, at some point $x\lesssim vt$. If a particle is that far on $x$ axis, 
it means that this particle
had spent at most time $t_y=t-x/v$ for its walks along the $y$ direction. 
Therefore, if we look at the PDF along the $y$ direction, 
it has been 'built' by only those particles which spend time $t_y$ evolving along 
this direction  (note that for the product model the time of walks 
along both directions is the same $t$ as both walks are independent). 
Therefore, for a given moment of time, the PDF along $x$-axis of the $XY$ model 
is not the 1D L\'{e}vy distribution uniformly scaled
with the pre-factor $1/t^{1/\gamma}$ (as in the product model) but a product of the L\'{e}vy distribution and the non-homogeneous factor  $t^{1/\gamma}/(t-x/v)^{1/\gamma}$; 
this pre-factor diverges as the particle gets closer to the front (see Fig. 3a) but it is integrable, so that the total number of particles is still conserved. 

\subsection{Probability density functions of 2D L\'{e}vy walks in the ballistic regime} For the product model, the PDF is given 
by the product of two PDFs of the one-dimensional L\'{e}vy walk. 
In the ballistic regime, PDF of the L\'{e}vy walk is the Lamperti distribution [S2]. 
For some particular values of $\gamma$ it has a compact form. 
For example, for $\gamma=1/2$ we have $P_{\text{LW}}^{(\gamma=1/2)}(x,t)=\pi^{-1}(v_0^2t^2-x^2)^{-1/2}$, 
and therefore, for 2D we get
\begin{equation}
P_{\text{prod}}^{(\gamma=1/2)}(x,y,t)=\frac{1}{\pi^2(v_0^2t^2-x^2)^{1/2}(v_0^2t^2-y^2)^{1/2}}.
\end{equation}
For the $XY$-model, the asymptotic expression for the propagator in the ballistic regime $0<\gamma<1$ is
\begin{equation}
P_{XY}(k_x,k_y,s)=\frac{(s+ik_xv_0)^{\gamma-1}+(s-ik_xv_0)^{\gamma-1}+(s+ik_yv_0)^{\gamma-1}+(s-ik_yv_0)^{\gamma-1}}{(s+ik_xv_0)^{\gamma}+(s-ik_xv_0)^{\gamma}+(s+ik_yv_0)^{\gamma}+(s-ik_yv_0)^{\gamma}}.
\end{equation}
For the uniform  model, the answer is more compact:
\begin{equation}
P_{\text{uniform}}(\mathbf{k},s)=\frac{\int_{0}^{2\pi}(s+ikv_0\cos\varphi)^{\gamma-1}\mathrm{d}\varphi}{\int_{0}^{2\pi}(s+ikv_0\cos\varphi)^{\gamma}\mathrm{d}\varphi},
\end{equation}
where $k=|\mathbf{k}|$. 
The integrals over the angle $\varphi$ can 
be calculated yielding hyper-geometric functions; 
the remaining technical difficulty is to calculate the inverse transforms. 
Recently, it was shown in the one-dimensional case, 
that the propagators of the ballistic regime can be 
calculated without explicitly performing the inverse transforms [S2]. 
This approach has been then generalized to 2D case for the uniform model [S3], where the density of particles was shown to be described by the following expression:
\begin{equation}
\label{cartesian}
P(\mathbf{r},t)_{\text{uniform}}=-\frac{1}{\pi^{3/2}t^2} D_-^{1/2}\Big\{\frac{1}{x^{1/2}}\operatorname{Im}\frac{_{2}F_1((1 - \gamma)/2, 1 - \gamma/2; 1; \frac{1}{x})}{_2F_1(-\gamma/2, (1- \gamma)/2; 1; \frac{1}{x})}\Big\}\left(\frac{\mathbf{r}^2}{t^2}\right)
\end{equation}
where $D_-^{1/2}$ is the right-side Riemann-Liouville fractional
derivative of order 1/2 (see [S3] for further details). It would be interesting to try to extend this formalism to random walks without rotational symmetry, such as for example the $XY$ model. The PDFs for the three models in the ballistic regime are shown in Fig.~S1.


\begin{figure}[t]
\includegraphics[width=0.99\textwidth]{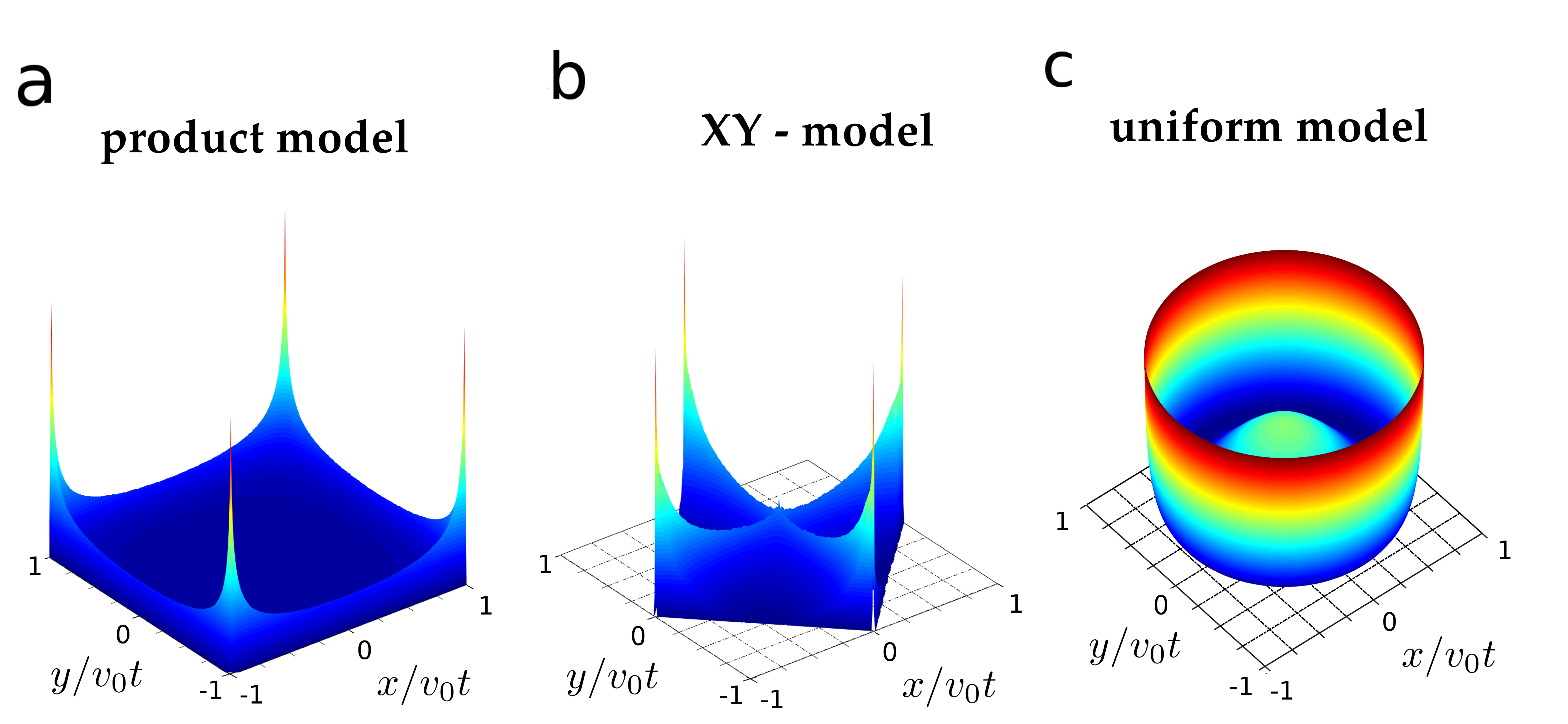}
\caption{\textbf{Probability density functions of the three models in the ballistic 
regime.}  The PDFs for the (a) product model, (b) $XY$-model, and (c) uniform model are plotted
on a log scale (color bars are not shown) for the time $t/\tau_0 = 10^4$.
The parameters are $\gamma = 1/2$, $\tau_0=v_0=1$.}
\label{FigS1}
\end{figure}



\begin{figure}[t]
\center
\includegraphics[width=0.85\textwidth]{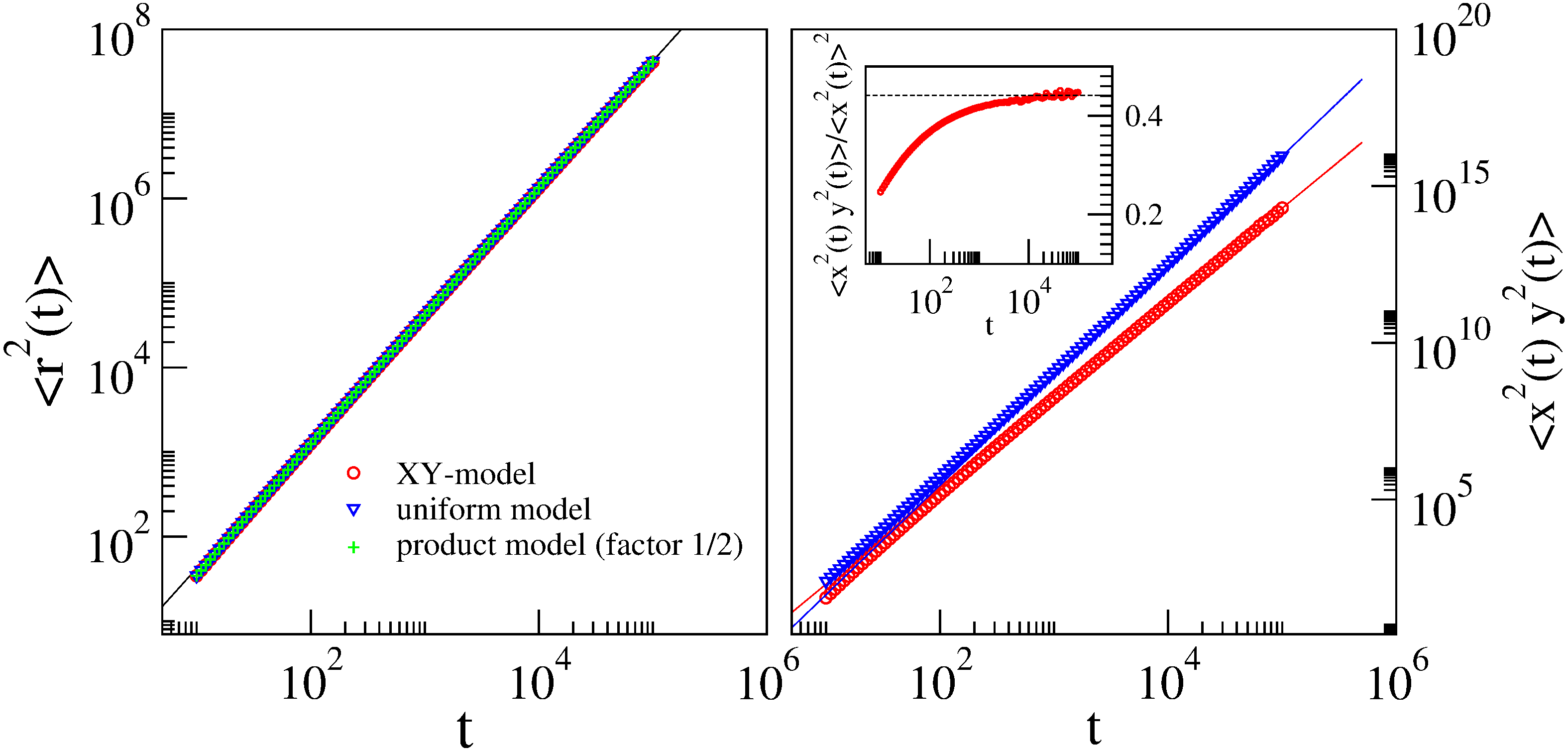}
\caption{\textbf{Second and forth moments of the three models in the super-diffusive regime, $\gamma=3/2$.} 
(left panel) The MSD is identical for all three models (though the product model needs an additional prefactor of 1/2). 
(right panel) The difference between the models becomes apparent in the cross-moment $\langle x^2y^2\rangle$. 
It scales as $t^{6-2\gamma}$ for the product and $XY$-model and ad $t^{5-\gamma}$ for the uniform model.
The inset shows that the convergence of the ration between the moments to the analytical $PC$ value (see Table S1)
in course of time (number of realizations is $10^5$ at every time point).
The parameters are $\tau_0=v_0=1$.}
\label{figS3}
\end{figure}

\subsection{MSD and higher moments} 
In case of an unbiased random walk that starts at zero, the MSD is defined as
\begin{equation}
\left<\mathbf{r}^2(t)\right>=\int{\rm d}\mathbf{r}\,\,\mathbf{r}^2 P(\mathbf{r},t),
\label{msd_def}
\end{equation}
The asymptotic behavior of the MSD can be 
calculated by differentiating the PDF in 
Fourier-Laplace space twice with respect to $\mathbf{k}$ and afterwards setting $\mathbf{k}=0$. 
\begin{equation}
\left<\mathbf{r}^2(t)\right>=-\nabla_{\mathbf{k}}^2\left.P(\mathbf{k},t)\right|_{\mathbf{k}=0}.
\label{msd_trick}
\end{equation}
In fact we can see from Eq.~\eqref{PksSI} 
that $\nabla_{\mathbf{k}}$ applied to $P(\mathbf{k},s)$ 
is equivalent to $-i\mathbf{v}\frac{{\rm d}}{{\rm d} s}$. 
Therefore all terms with the first order derivatives disappear 
due to the integration with a symmetric velocity distribution. 
Only the terms with second derivatives contribute to the answer:
\begin{equation}
\left.-\nabla_{\mathbf{k}}^2P(\mathbf{k},s)\right|_{\mathbf{k}=0}=\int{\rm d}\mathbf{v}\,\mathbf{v}^2h(\mathbf{v})\left\{\frac{\Psi(s)\cdot\psi''(s)}{[1-\psi(s)]^2}+\frac{\Psi''(s)}{1-\psi(s)}\right\}
\end{equation} 
Now, to calculate the scaling of the MSD in real time for 
different regimes of diffusion, we need 
to take the corresponding expansions of $\psi(s)$ and $\Psi(s)$ for small $s$, Eq.~\eqref{expansion}, 
and perform the inverse Laplace transform. As a result we obtain
\begin{equation}
\left<\mathbf{r}^2(t)\right>=\left\{
                \begin{array}{cc}
                 \vspace{5pt} v_0^2(1-\gamma)t^2 & 0<\gamma<1\\
                \vspace{5pt} \frac{2v_0^2\tau_0^{\gamma-1}(\gamma-1)}{(3-\gamma)(2-\gamma)} t^{3-\gamma}&1<\gamma<2\\
                  \frac{2v_0^2\tau_0}{\gamma-2}t & \gamma>2
                \end{array}
              \right.
              \label{msd_scaling}
\end{equation}
It is remarkable that for $0<\gamma<1$ the scaling exponent of the MSD is independent of the tail 
exponent $\gamma$ of the flight time distribution. An important observation is that the models 
are indistinguishable on the basis of their MSD behavior, see Fig S2 (left panel). 
Similarly, we  compute the forth moment and 
find that they scale similarly for all three models (though the prefactors are model-specific, see second column in 
Table S1). 
The difference between the models 
become tangible in the cross-moments $\langle x^2y^2\rangle$. 
We use these moments to define a generalized Pearson coefficient which is denoted by $PC$,
\begin{equation}
PC(t)=\frac{ \langle x^2(t)y^2(t)\rangle}{ \langle x^2(t)\rangle \langle y^2(t)\rangle}.
\end{equation}
We concentrate on the sub-ballistic super-diffusive regime and 
summarize the corresponding results in  Table S1. 

\begin{table*}
\renewcommand{\arraystretch}{2.5}
\begin{tabular}{|c|c|c|c|c|}
\hline
 Moment  & $\langle r^2\rangle$ & $\langle x^4\rangle=\langle y^4\rangle$    &     $\langle x^2y^2\rangle$        & $PC$                               \\[0.20cm] 
  \hline
Product          & $\frac{4v_0^2\tau_0^{\gamma-1} (\gamma-1)}{(2-\gamma)(3-\gamma)}t^{3-\gamma}$  & $\frac{4v_0^4 \tau_0^{\gamma-1}(\gamma-1)}{(4-\gamma)(5-\gamma)}t^{5-\gamma}$ &    $\frac{4v_0^4 \tau_0^{2\gamma-2}(\gamma-1)^2}{(2-\gamma)^2(3-\gamma)^2}t^{6-2\gamma}$ & 1  \\ [0.20cm] 
$XY$             & $\frac{2v_0^2\tau_0^{\gamma-1} (\gamma-1)}{(2-\gamma)(3-\gamma)}t^{3-\gamma}$  & $\frac{2v_0^4\tau_0^{\gamma-1}(\gamma-1)}{(4-\gamma)(5-\gamma)}t^{5-\gamma}$ &    $\frac{v_0^4\tau_0^{2\gamma-2}\gamma(\gamma-1)^4\Gamma^2[1-\gamma]}{\Gamma[7-2\gamma]}t^{6-2\gamma}$  & $\frac{\gamma\Gamma[4-\gamma]^2}{\Gamma[7-2\gamma]}$     \\ [0.20cm]
Uniform          & $\frac{2v_0^2\tau_0^{\gamma-1} (\gamma-1)}{(2-\gamma)(3-\gamma)}t^{3-\gamma}$  & $\frac{3v_0^4\tau_0^{\gamma-1}(\gamma-1)}{2(4-\gamma)(5-\gamma)}t^{5-\gamma}$ &   
  $\frac{v_0^4\tau_0^{\gamma-1} (\gamma-1)}{2(4-\gamma)(5-\gamma)}t^{5-\gamma}$ & $\frac{(\gamma -3)^2 (\gamma -2)^2}{2 (5-\gamma) (4-\gamma) (\gamma -1)}(\frac{t}{\tau_0})^{\gamma -1}$  \\[0.20cm]  \hline%
\end{tabular}
\caption{Asymptotic moments of the  models.}\label{table} 
\end{table*}

\begin{figure}[t]
\includegraphics[width=0.85\textwidth]{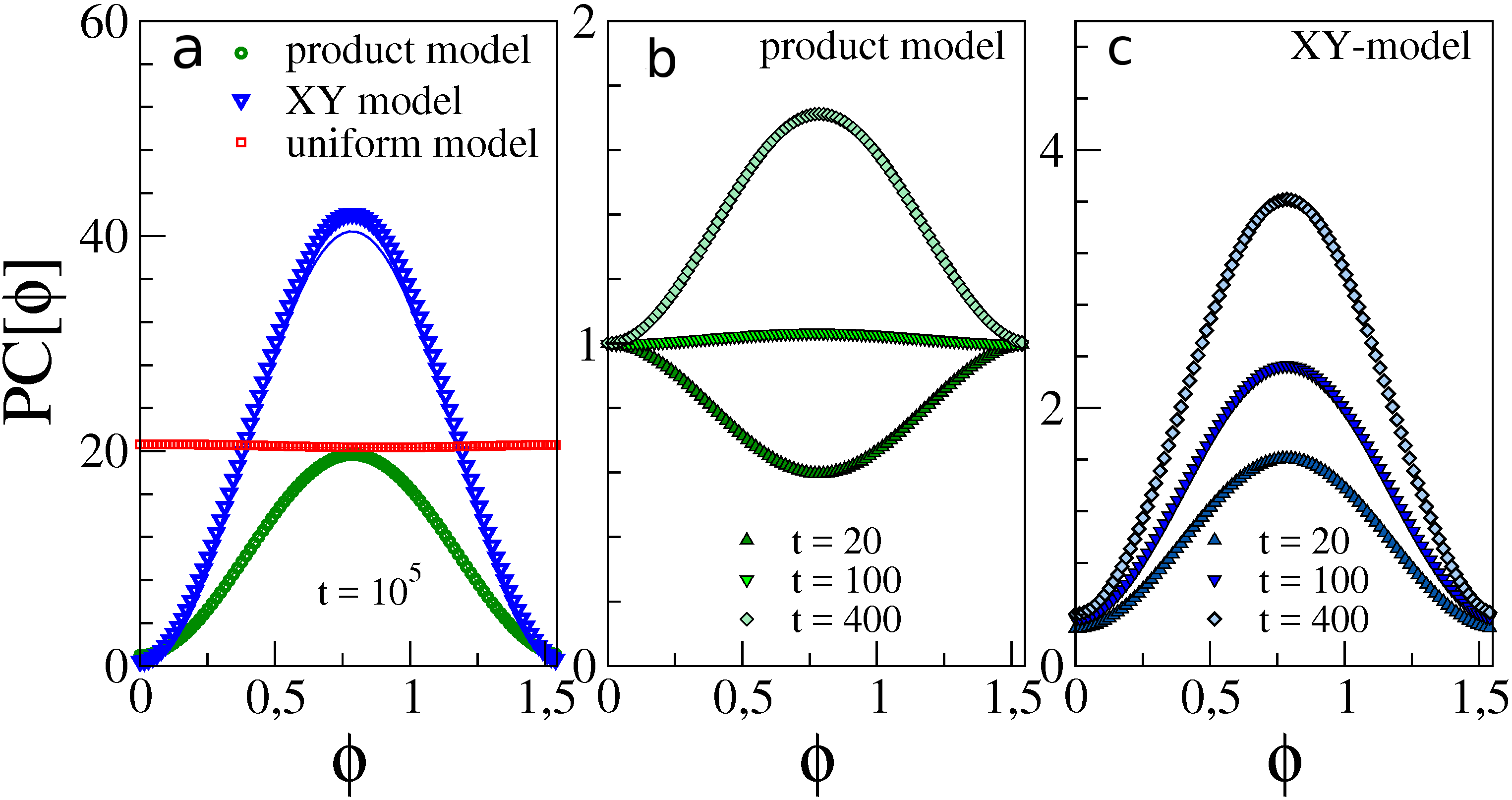}
\caption{\textbf{Frame dependence and time evolution of the Pearson coefficient.} 
a) Dependence of the Pearson coefficient on the frame orientation angle $\phi$
for the three models in the long-time limit. Symbols correspond to the results of the statistical sampling for  $t=10^5$
and lines present analytical results for the asymptotic limit, Eq.~S19. 
b-c) Time dependence of $PC[\phi]$ for the product and $XY$ models. For the short times, $t \lesssim 10^2$,
minima of the Pearson coefficient for the product model correspond to the diagonal and anti-diagonal directions while 
for larger  times, $t > 10^2$, these directions correspond to the maxima;  
minima are attained along the $\{x,y\}$ axes, see Fig.~1d in the main text. 
For the $XY$ model the positions of maxima and minima are time-independent. 
The parameters are the same as in Fig.~2 in the main text. Note that the mean time of ballistic flights is $\langle\tau\rangle = 2$
for the chosen set of parameters.
}
\label{FigS3}
\end{figure}

\subsection{Angular dependence of the Pearson coefficient} 
As mentioned in the main text, the value of $PC$ depends on the choice of the coordinate frame. 
To quantify the dependence of the $PC$ on the orientation of the frame of references we consider 
random walks $\{x'(t),y'(t)\}$ which are obtained by turning three original random walks 
(uniform, product and $XY$) by an arbitrary angle $0\leq\phi\leq\pi/2$ with respect to the $x$-axis and calculate the 
corresponding $PC[\phi]$. The coordinates in the turned and original frames are related via:
\begin{equation}
\{x'(t),y'(t)\}=\{x(t)\cos\phi-y(t)\sin\phi, x(t)\sin\phi+y(t)\cos\phi)\}
\end{equation} 
After simple algebra and by discarding terms with odd powers 
of $x$ and $y$, which will disappear after averaging due to symmetry, we obtain:
\begin{equation}
PC[\phi]=\frac{\langle x^2y^2\rangle(\cos^4\phi+\sin^4\phi-\sin^22\phi)+(\langle x^4\rangle+\langle y^4\rangle)\cos^2\phi\sin^2\phi}{\langle x^2y^2\rangle(\cos^4\phi+\sin^4\phi)+(\langle x^2\rangle^2+\langle y^2\rangle^2)\cos^2\phi\sin^2\phi}.\label{pcphi}\end{equation}
This expression can now be 
evaluated by using the fourth and second-order 
cross moments of the original random walks given in Table S1. 
It can be shown that for the product and $XY$ models and for any $\phi\neq \{0,\pi/2\}$, $PC(t;\phi)\propto t^{\gamma-1}$, 
whereas for the uniform model $PC(t;\phi)\propto t^{\gamma-1}$ is naturally independent of the angle $\phi$.
The angle dependent Pearson coefficients are different for all three models, see Fig. S3a. 
Importantly, the Pearson coefficient contains only one ``unknown'' parameter $\tau_0$, 
the scaling parameter of the flight time PDF (recall that the exponent $\gamma$ of the power 
law tail in the flight time distribution can be determined from the scaling of the MSD). 
Thus by measuring the $PC(t;\phi)$, for example, at two different time points would allow for a unambiguous 
determination of $\tau_0$ and of the type of the random walk model.

An interesting observation can be made by numerically calculating $PC[\phi]$ as a function of time. 
While  the shape of the dependence remains the same for the uniform- (flat at any time) and $XY$-model (see Fig.S3c), 
for the product model it changes qualitatively. The minima of the $PC_{\text{prod}}(t;\phi)$ 
for short times transform into the maxima for large times passing through an intermediate flat shape, 
see Fig.~S3b. This is a quantification of the transition we also observe by sampling individual trajectories. It 
is a consequence of the fact that for short times the preferred direction of motion are diagonals and anti-diagonals (inset in Fig.1d), whereas 
for large times the trajectories resemble those of the $XY$-model (Fig. 1d,e). Correspondingly the maximum of 
the $PC[\phi]$ ``shifts'' by $\pi/4$ as time progresses. At the same time, the positions of maxima and minima of 
$PC[\phi]$ for the $XY$ model are time-invariant. This effect can be used  to distinguish between the two models.

\subsection{Numerical sampling procedure} The statistical sampling of the model PDFs was performed
on the GPU clusters of the Max Planck Institute for the Physics of Complex Systems (Dresden) 
(six NVIDIA M2050 cards) and the university of Augsburg (four TESLAK20XM cards). Together with the
straightforward parallelization of the sampling process this allowed us to collect up to $10^{14}$
realizations for every set of parameters.

\subsection{Hamiltonian diffusion in an egg-crate potential} We use model from Refs.~[S4,S5]. The model Hamiltonian
has the form
\begin{eqnarray}
H(x,y,p_x,p_y)&=& E = \frac{p_x^2}{2}+\frac{p_y^2}{2}+A\\ &+&B(\cos x +\cos y)+ C\cos x \cos y, 
\end{eqnarray}
with particular value of the parameters $A=2.5$ (this parameter can be dropped since it 
does not enter the corresponding equations of motion; however, we preserve the original form of 
the Hamiltonian given in Refs.~[S4,S5]), $B=1.5$, and $C=0.5$. To integrate the equations of motion, 
we use the
high-order symplectic integrator $SBAB_5$ [S6] with time step $\triangle t = 10^{-3}$. The energy 
of the system was conserved as $|\triangle E/E| < 10^{-10}$ during the whole integration time. 

A typical trajectory of the system is shown on Fig.~S4 (left panel). 
The PDF of the corresponding dispersal for time $t = 10^3$
was sampled with $10^8$  realizations; the obtained result is shown in Fig.~S4 (right panel).
Figure S5 presents the results of numerical simulations performed to 
estimate Pearson coefficient $PC$ of the dispersals
for two different values of energy $E$.

\begin{figure}[t]
\center
\includegraphics[width=0.95\textwidth]{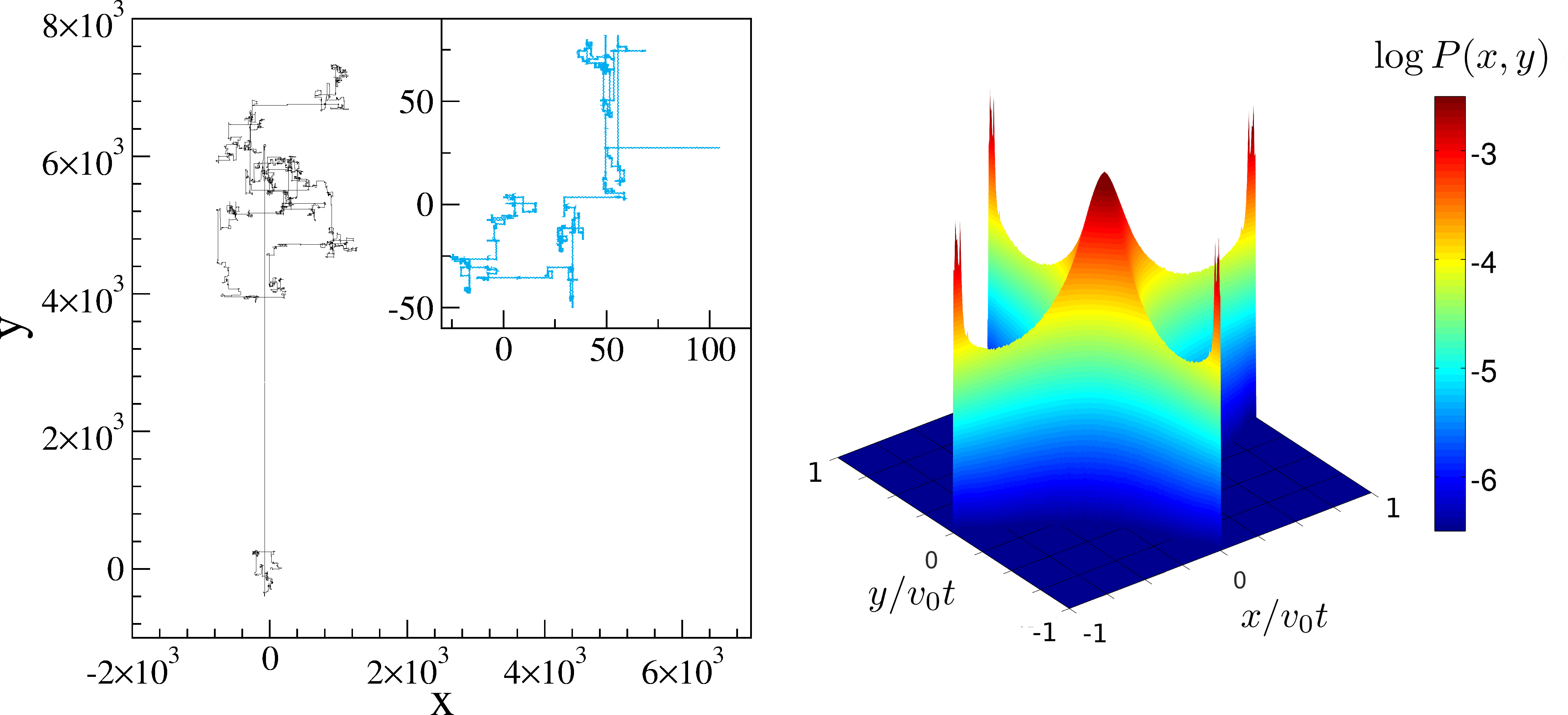}
\caption{\textbf{Hamiltonian particle diffusion in the egg-crate potential, Eq.~(S25).} (left panel) 
A typical trajectory for $E=4$ produced by the particle after time $t = 10^5$. 
Inset shows a trajectory for  time $t = 10^3$; (right panel) The PDF of the corresponding 
dispersal process for $t = 10^3$
sampled with $10^8$ realizations.}
\label{figS4}
\end{figure}

\begin{figure}[b]
\center
\includegraphics[width=0.75\textwidth]{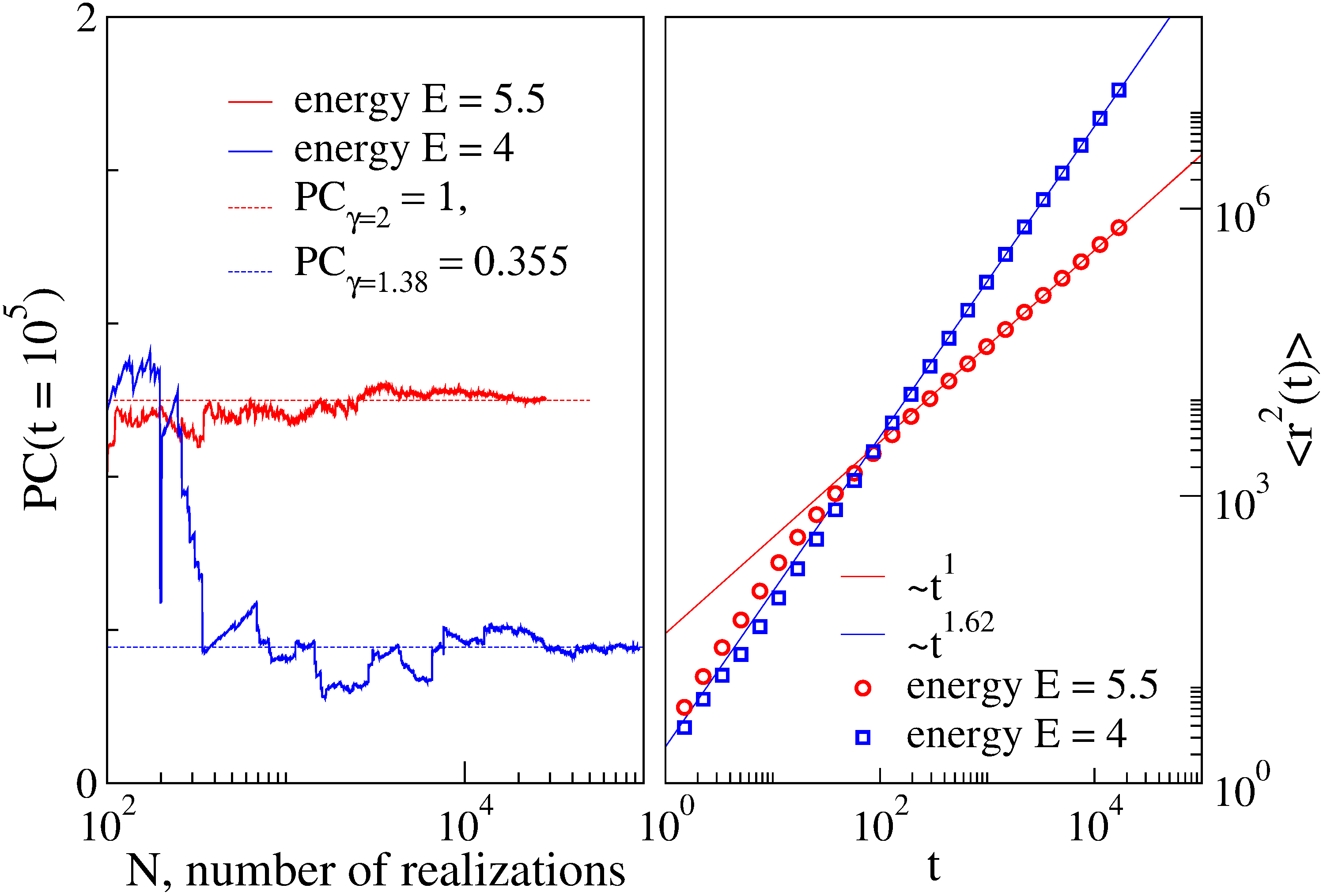}
\caption{\textbf{Pearson coefficient and moment scaling for the Hamiltonian dispersal.}  $PC$s of the dispersals 
for two energy values were sampled (left panel). Plot shows Pearson coefficients for the fixed time $t = 10^5$ as a function of 
number of collected realizations. Thin dashed lines correspond to the analytical results, Tabe S1, obtained 
for exponents $\gamma$ extracted from the MSD of the dispersals (right panel).}
\label{figS5}
\end{figure}


\vspace*{12pt}
\noindent\textbf{\large{{References}}}

\noindent [S1] A. Erd\'{e}lyi, Tables of Integral Transforms, Vol. 1 (McGraw-Hill, New York, 1954)

\noindent [S2] D. Froemberg, M. Schmiedeberg, E. Barkai, and V. Zaburdaev, Asymptotic densities of 
ballistic L\'{e}vy walks. Phys. Rev. E \textbf{91}, 022131 (2015).

\noindent [S3] M. Magdziarz, and T. Zorawik, Explicit densities of multidimensional ballistic L\'{e}vy walks, Phys. Rev. E \textbf{94}, 022130 (2016).

\noindent [S4] T. Geisel, A. Zacherl, and G. Radons, 
Generic $1/f$ noise in chaotic Hamiltonian dynamics. Phys. Rev. Lett. \textbf{59}, 2503--2506 (1987).

\noindent [S5] J. Klafter and G. Zumofen, L\'{e}vy statistics in a Hamiltonian system. Phys. Rev. \textbf{E} 49,
4873--4877 (1994).

\noindent [S6] J. Laskar, and P. Robutel, High order sympletic integrators for perturbed Hamiltonian systems.
Celest. Mech. Dyn. Astron. \textbf{80}, 39 (2001).
\end{widetext}

\end{document}